\documentclass[fleqn,10pt]{wlscirep}

\usepackage{amsmath}

\usepackage{amsfonts}

\usepackage{graphicx}

\usepackage{textcomp}

\usepackage[T1]{fontenc}	

\usepackage{lmodern}

\usepackage[utf8]{inputenc} 

\usepackage[english]{babel}

\usepackage{physics}
\usepackage[nameinlink]{cleveref}

\usepackage{mathptmx}

 \newcommand{\bla}[1]{\left(#1\right)}
 \newcommand{\blb}[1]{\left[#1\right]}

\begin{document}

\title{NV center based nano-NMR enhanced by deep learning}

\author[1,*]{Nati Aharon}
\author[1]{Amit Rotem}
\author[2]{Liam P. McGuinness}
\author[2]{Fedor Jelezko}
\author[1]{Alex Retzker}
\author[1]{Zohar Ringel}

\affil[1]{Racah Institute of Physics, The Hebrew University of Jerusalem, Jerusalem 
91904, Givat Ram, Israel}
\affil[2]{Institute for Quantum Optics, Ulm University, Albert-Einstein-Allee 11, Ulm 89081, Germany}
\affil[*]{nati.aharon@mail.huji.ac.il}

\begin{abstract}
The growing field of nano nuclear magnetic resonance (nano-NMR) seeks to estimate spectra or discriminate between spectra of minuscule amounts of complex molecules. While this field holds great promise, nano-NMR experiments suffer from detrimental inherent noise.
This strong noise masks to the weak signal and results in a very low signal-to-noise ratio.
Moreover, the noise model is usually complex and unknown, which renders the data processing of the measurement results very complicated. Hence, spectra discrimination is hard to achieve and in particular, it is difficult to reach the optimal discrimination. In this work we present strong indications that this difficulty can be overcome by deep learning (DL) algorithms. The DL algorithms can mitigate the adversarial effects of the noise efficiently
by effectively learning the noise model.
We show that in the case of frequency discrimination DL algorithms reach the optimal discrimination without having any pre-knowledge of the physical model. Moreover, the DL  discrimination scheme outperform Bayesian methods when verified on noisy experimental data obtained by a single Nitrogen-Vacancy (NV) center.  In the case of frequency resolution we show that this approach outperforms Bayesian methods even when the latter have full pre-knowledge of the noise model and the former has none. These DL algorithms also emerge as much more efficient in terms of computational resources and run times.  
Since in many real-world scenarios  the noise is complex and difficult to model, we argue that DL is likely to become a dominant tool in the field.
\end{abstract}
\maketitle

\section*{Introduction}
The newly developed discipline of nano-NMR \cite{balasubramanian2008nanoscale,gruber1997scanning,maze2008nanoscale,staudacher2013nuclear,mamin2013nanoscale,muller2014nuclear,devience2015nanoscale} is aimed at reducing the minimal NMR sample size by many orders of magnitude, and thus increasing the NMR sensitivity and spatial resolution down to a few molecules \cite{lovchinsky2016nuclear}. This is achieved by replacing the macroscopic coil of the NMR setup, which measures the magnetic field, by a single or an ensemble of controllable spins, e.g., NV centers in diamond, which serve as tiny magnetometers.
Recent experiments have shown that it is possible to estimate the spectrum of artificial signals and signals of polarized samples with high resolution \cite{schmitt2017submillihertz,boss2017quantum,bucher2017high,rotem2017limits,zaiser2016enhancing}. However, the obvious advantages of receiving spectral information about tiny quantities of molecules are masked by the extra amount of noise that goes hand in hand with most configurations of this setup. There are a few sources of this extra noise, which include the NV coherence time (magnetic noise), the controller noise (laser and microwave operations), and most importantly the diffusion induced noise, which is negligible in the regular NMR setup but is extremely large in the nano-NMR setup and broadens the line-width above the required resolution.
This noise  creates a serious bottleneck, as the crucial information is encoded in the tiny chemical shifts and small energy gaps caused by J - couplings. That is, the nano-NMR setup is usually characterized by a weak measured signal, which is masked by a strong noise.  

Moreover, the precise noise model is usually complex and unknown.  Consequently, it is an intractable data processing challenge to achieve a spectral discrimination between weak and similar signals of near-by frequencies. In particular, because the noise model is complex and unknown,  it is difficult to tackle this noise and reach the optimal discrimination by conventional data analysis methods with which optimal discrimination can usually be obtained only when full knowledge of the noise model is available.     

In this work we show that the challenge of spectral discrimination between weak and similar signals in the presence of strong and complex noise, can be efficiently confronted by DL algorithms, which effectively learn the noise model.
Moreover, we show that DL methods are capable of learning the noise model from a small amount of data which only needs to be gathered for a few minutes. This means that a DL algorithm can analyze a test signal with the same efficiency as numerically demanding Bayesian methods that rely on precise knowledge of the model.  In addition, we show that DL methods can be extremely useful in dealing with challenging frequency resolution problems and possibly overcome Bayesian methods even under  assumptions that these have  full knowledge of the model and infinite computing power.

\begin{figure}[t]
\begin{center}
\includegraphics[width=0.6\textwidth]{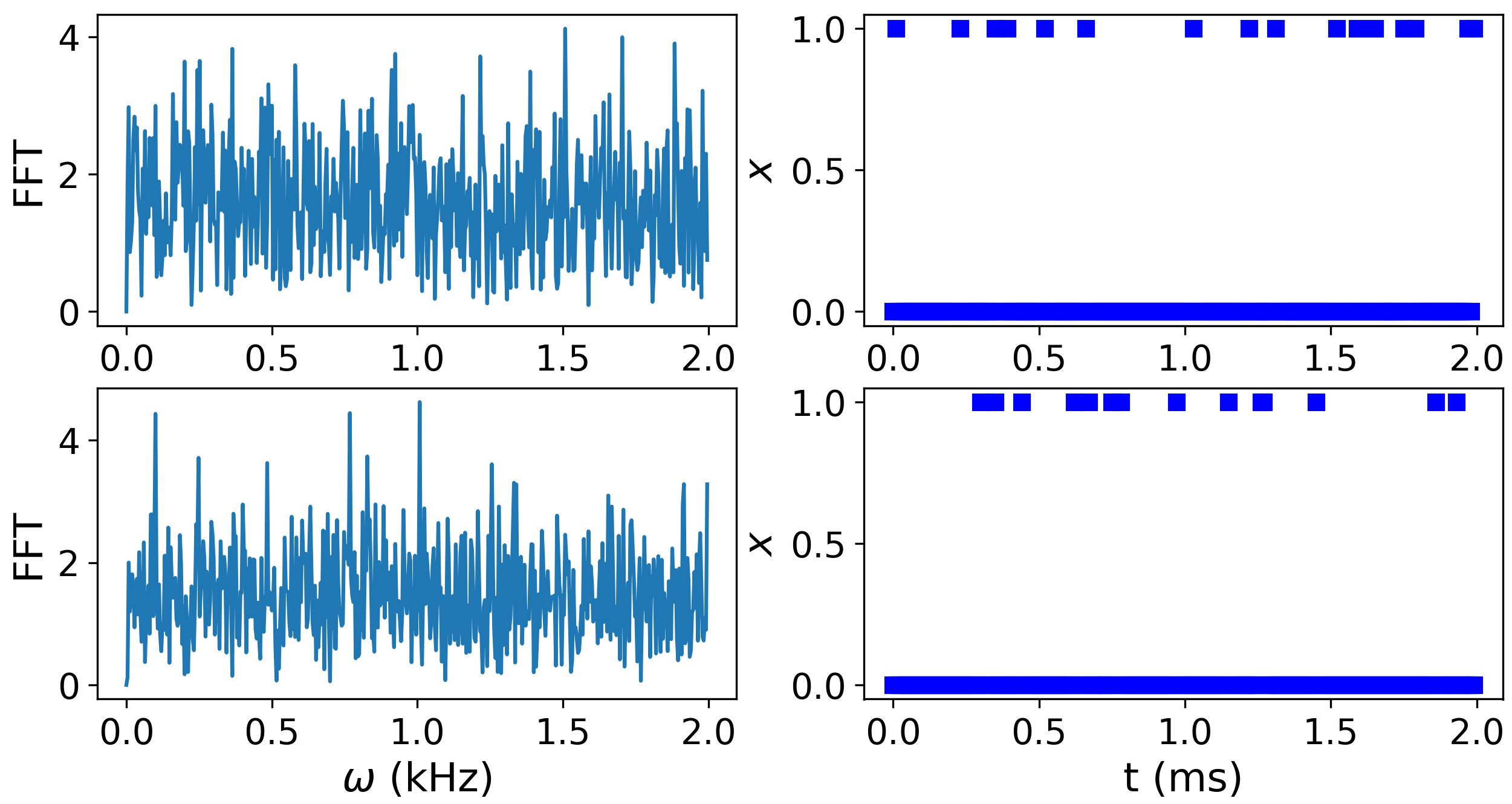}
\end{center}
\caption{Typical noisy data of two different frequencies that we aim to discriminate in this work.  The oscillating magnetic signals at the two different frequencies suffer from a strong phase noise and are read by an NV center, which adds quantum noise to the output binary signal (see Eq. \ref{probloweff}).
(upper right): The time trace binary signal from one frequency of $250$ Hz together with its Fourier transform after subtracting the zero frequency (upper left). (lower right): The time trace binary signal from the second frequency of $251.6$ Hz and its Fourier transform (lower left). }
\label{two_freq}
\end{figure}

DL techniques have been successfully applied to spectral data in the fields of Astronomy, Chemistry, Geosciences, and Bioinformatics  \cite{ReviewSpectralLearning2008}. Spectral data from  these disciplines pose similar challenges: (1) High data dimensionality (2) Difficulty of modeling the important features from first principles (3) Dirty environments with many classes of objects that need to be differentiated along with varying signal intensities (4) Importance of subtle differences in the signal. Despite these difficulties, which apply in our context as well,  impressive achievements have been made such as the detection of narcotics in Raman spectroscopy data with a 0.5\% error rate \cite{Howley2006}. 

DL methods have also been used for the analisys of NMR data, in particular in the context of automated protein structure for peak-picking of nuclear magnetic resonance spectra \cite{carrara1993neural},  of biological macromolecules \cite{corne1992artificial}, and recently also in the context of analyzing a variety of spectral images of proteins by using support vector machine classifier combined with histogram of oriented gradient \cite{klukowski2015computer}, and by using convolutional neural network \cite{klukowski2018nmrnet}. In addition, deep learning techniques such as long short term memory networks\cite{Li1} and variational auto-encoder networks \cite{Li2} have been used in NMR applications for material characterization and subsurface characterization

We believe that the success of DL methods in the analysis of regular NMR data  should be amplified in the nano-NMR setup due to the larger amount of noise in this setting, which originates from two main ingredients that are absent in the regular NMR setting. The first ingredient is the origin of the signal. While in regular NMR the signal is created by thermal polarization, in nano-NMR the signal is created by statistical polarization \cite{Herzog2014}, which imply that in the nano-NMR setup the noise is stronger. The second ingredient is the quantum projection noise, which is of a Poissonian or a Bernouli nature, and in many cases is the dominant source of noise. Here we provide evidence that DL methods can tackle these noises efficiently. 

\begin{figure}[t]
\begin{center}
\includegraphics[width=0.6\textwidth]{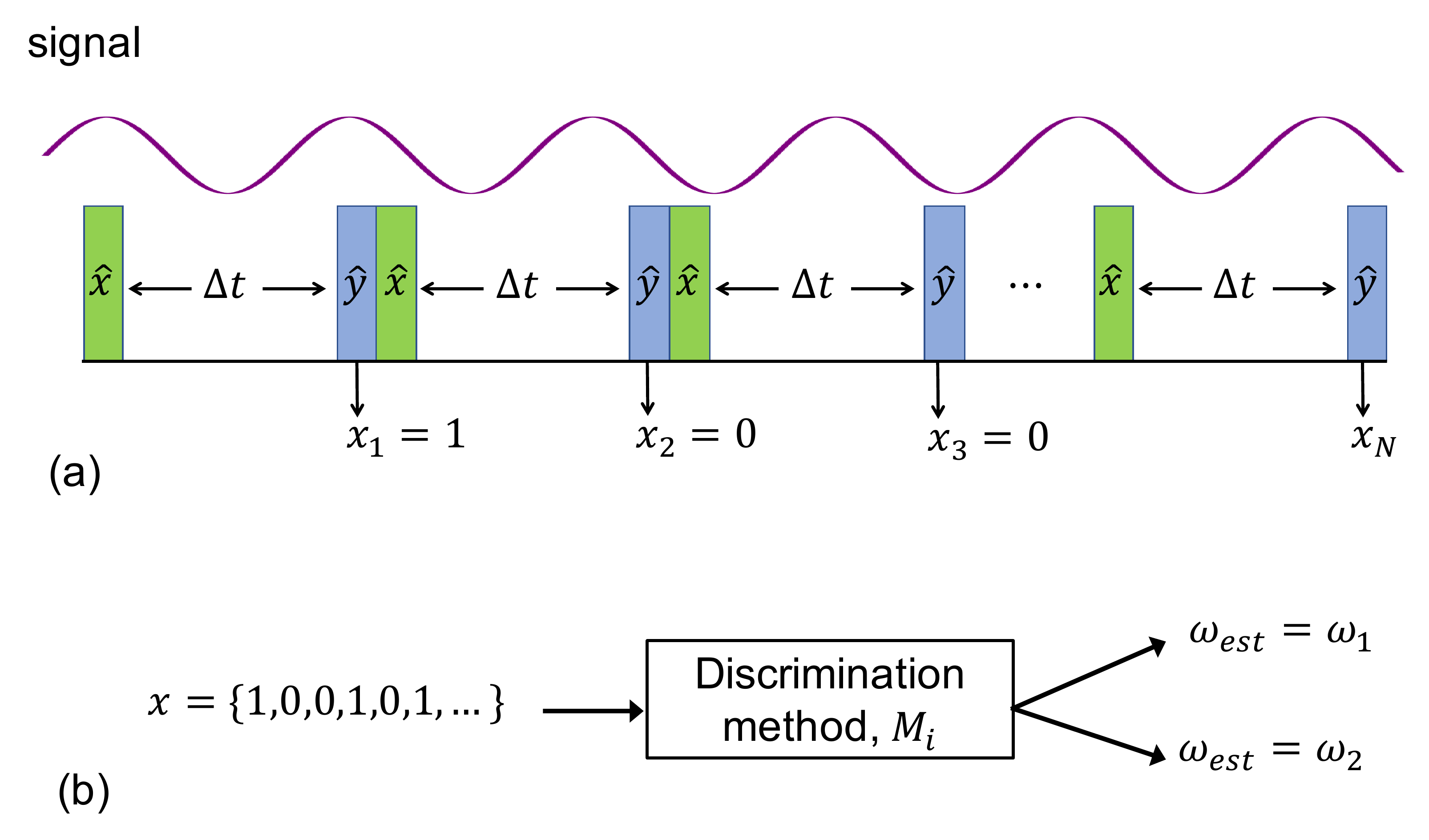}
\end{center}
\caption{The physical model. (a) The probe, which is initially polarized along $\hat{x}$, freely evolves according to $H_{s_i}$ (Eq. \ref{eq_single_signal}) for a short  duration, $\Delta t$, and then is measured along $\hat{y}$. In the measurement scheme of a single experiment, the sequence of probe operations consists of initialization, evolution, and measurement, which is repeated $N$ times under the constant presence of a signal. In each experiment, the frequency of the signal is equal to one of two known frequencies, $\omega_1$ and $\omega_2$. (b) Our aim is to discriminate between the two frequencies. A single experiment results in a string of bits, $x = \{1,0,0,1,...\}$. Given $x$, we want to obtain an estimation of the frequency of the signal, $\omega_{est}=\omega_1$ or $\omega_{est}=\omega_2$.}
\label{model}
\end{figure}

To evaluate the efficiency of DL methods in terms of the spectroscopy of nano-NMR data, we consider two problems, frequency discrimination and frequency resolution. We first examine the ability of DL methods to discriminate between two signals corresponding to two different frequencies. In particular, we consider data from signals that were read by an NV center, which simulates noisy nano-NMR data. Typical data for these two frequencies are shown in Fig. \ref{two_freq}, which presents two time traces of the datasets together with their Fourier transforms. It is  immediately clear that it is impossible to discriminate between the two frequencies based on the Fourier transform alone because the signal has a strong phase noise on top of the detection noise. 
In this work we show that DL methods 
are able to classify the data with the same efficiency as  Bayesian methods, which use full knowledge of the signal and noise model and are numerically much more demanding than DL methods. 
The advantage of DL methods is also indicated by their superior performance in  frequency discrimination of the experimental data, where the signal and noise models are not fully known.   
 
We then employ DL methods to tackle the problem of frequency resolution in a noisy environment. We show that DL methods can efficiently discriminate between  the signal of a single frequency and the signal of two nearby frequencies that have a strong amplitude and phase noise.

Our results strongly suggest that DL methods can effectively learn the physical and noise models and by that constitute an efficient alternative to Bayesian methods, which require a priori knowledge on the physical and noise models.

\begin{figure}[t]
\begin{center}
\includegraphics[width=0.57\textwidth]{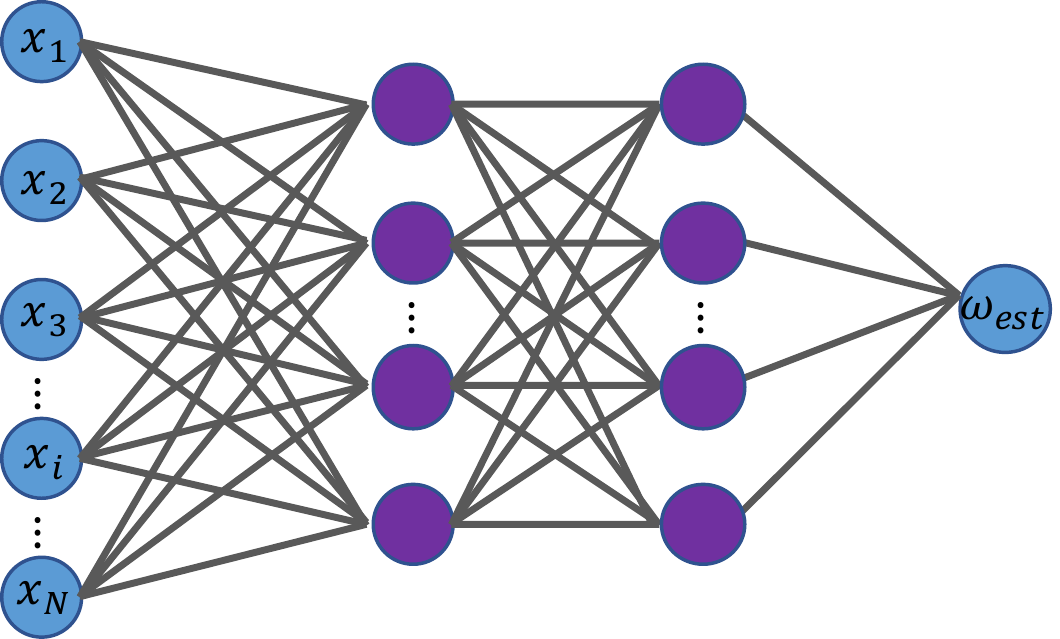}
\end{center}
\caption{The $M_{DL}$ neural network is a feed-forward fully connected neural network. The input layer inputs the measurement results $x$ to the second layer (first hidden layer). The output of the last hidden layer is fed to the output layer, which results in the frequency discrimination, $\omega_{est}=\omega_1$ or $\omega_{est}=\omega_2$.}
\label{network}
\end{figure}

\section*{Frequency discrimination} 
\subsection*{The physical model}
We consider the problem of discrimination between two signals corresponding
to two different frequencies by a single quantum probe. In the nano-NMR setup this corresponds, for example, to the scenario where a single NV center, which serves as a tiny magnetometer, is placed in the proximity of a sample that contains two known molecules between which we wish to discriminate.      Specifically, in the presence of a single frequency signal (a single molecule) the Hamiltonian of the spin probe is given by
\begin{equation}
\label{eq_single_signal}
H_{s_i} = g_i \cos(\omega_i t + \phi_i) S_z,
\end{equation}
where $g_i$, $\omega_i$, and $\phi_i$ are the amplitude, frequency, and (random) phase of signal $i$ respectively, which is the standard setting in nano-NMR experiments    \cite{balasubramanian2008nanoscale,gruber1997scanning,maze2008nanoscale,staudacher2013nuclear,staudacher2015probing}. The probe, which is initially polarized along $\hat{x}$, freely evolves according to $H_{s_i}$ for a short  duration, $\Delta t$, and then is measured along $\hat{y}$. In the measurement scheme of a single experiment, the sequence of probe operations consists of initialization, evolution, and measurement, which is repeated many times under the constant presence of a signal (Fig. \ref{model} (a)). In the case of a single shot measurement, the measurement result is a sequence of zeros and ones, Fig. \ref{two_freq} (right), and the 
probability for a successful measurement (one) is given by
\begin{equation}
\label{probideal}
P(t)=\sin\blb{\frac{g_i}{2\omega_i}\bla{\sin\blb{\omega_i t + \phi_i}-\sin\blb{\omega_i \bla{t-\Delta t} + \phi_i}}+\frac{\pi}{4}}^2.
\end{equation}
We start by considering an ideal scenario (no noise or inefficiencies) where Eq. (\ref{probideal}) holds. We assume that in each experiment the signal corresponds to one of two known frequencies ($\omega_1$ and $\omega_2$), the amplitudes of the signals are known, but in each experiment the signal has an unknown uniformly distributed random phase. A single experiment results in a string of bits, $x = \{1,0,0,1,...\}$, where $1$ and $0$ correspond to a detection of the $m_s = 0$ state or $m_s = -1$ state of the NV center. Given $x$, we want to obtain an estimation of the frequency of the signal, $\omega_{est}=\omega_1$ or $\omega_{est}=\omega_2$ (Fig. \ref{model} (b)). We quantify the performance of a discrimination method $M$ by the error probability of the frequency estimation, which is defined by
\begin{equation}
\label{eq_error}
P^{error}_M \equiv \frac{1}{2} \sum_{i=1 \atop j\neq i}^{i=2} P_M(\omega_{est}=\omega_j \vert \omega_i),
\end{equation}
where $P_M(\omega_{est}=\omega_j \vert \omega_i)$ is the probability of method $M$ to output $\omega_{est}=\omega_j$ given that the frequency of the signal is $ \omega_i$.

\subsection*{Full Bayesian method}
In the ideal scenario considered here, we have full knowledge of the model (Eq. (\ref{probideal})) and the only unknowns are the random phases $\phi_i$. Hence, we can simply utilize a Full Bayesian method  known as the likelihood-ratio test and denoted by $M_{FB}$, where for each frequency we calculate the maximal log-likelihood over the unknown random phases. That is, 
\begin{equation}
L_1 = \max_{\phi_k} L(\phi_k|x,\omega_1),\quad  L_2 = \max_{\phi_k} L(\phi_k|x,\omega_2),
\end{equation}
where 
\begin{eqnarray}
 L(\phi_k|x,\omega_i) &=& \sum_j (x_j \log P(t_j,\omega_i,\phi_k)\nonumber\\
  &+& (1-x_j) \log (1- P(t_j,\omega_i,\phi_k))).
\end{eqnarray}
We estimate the frequency according to the larger likelihood; that is
\begin{equation}
\omega_{est} = \begin{cases}
 \omega_1 & L_1 > L_2\\
\omega_2 &\text{otherwise}.
\end{cases}
\end{equation}
As  $M_{FB}$ utilizes the maximal information on the signal, it obtains the minimal possible error for an unbiased estimator, which can serve as a benchmark to evaluate the efficiency of a learning method. Hence, its error probability serves as a lower bound for the DL method. It is known that Bayesian methods are optimal given the maximal amount of information and given that the optimization can be done efficiently, which is usually not the case, specifically when considering a noisy environment. In order to verify that  we indeed have the optimal method, we compare the results to an analytical calculation of the Fisher Information (FI), which can be done in this case. 

In general, full knowledge is not available due to either a lack of knowledge of the noise model in the experiment and detection inefficiencies, or lack of knowledge  of the signal. In this case, we can utilize a correlation based method, $M_{corr}$, for frequency discrimination. To this end, we first use a train set of measurement results, $X_{train}$, for which the frequency of the signal is known. For each $x\in X_{train}$ we calculate the correlation vector $C_{k}=  \langle x_i x_{i+k}\rangle_i $ (here we replace the $0$ bit by $-1$). Then, for each frequency we calculate the averaged correlation vector, $C^{\omega_i}= \langle C_{k} \rangle_{x\in X_{train}^{\omega_i}}$, where $X_{train}=X_{train}^{\omega_1}\cup X_{train}^{\omega_2}$. To estimate the frequency of an unknown signal we calculate its correlation vector, $C_{k}$, and then the distances
\begin{equation}
D_1 = ||C_{k}-C^{\omega_1}||_{L_2}, \quad D_2 = ||C_{k}-C^{\omega_2}||_{L_2},
\end{equation}
by the $L_2$ norm. We estimate the frequency according to the smaller distance; that is, 
\begin{equation}
\omega_{est} = \begin{cases}
 \omega_1 & D_1 < D_2\\
\omega_2 &\text{otherwise}.
\end{cases}
\end{equation}
This method, however, disregards higher order correlations functions and the finite precision of the correlation functions itself which varies considerably between the nearest neighbors and the higher neighbor separation. While in the limit where all these effects are taken into account this should approach the optimum, it is numerically very challenging or even impossible to apply to most problems of interest.

\begin{figure}[t]
\begin{center}
\includegraphics[width=1\textwidth]{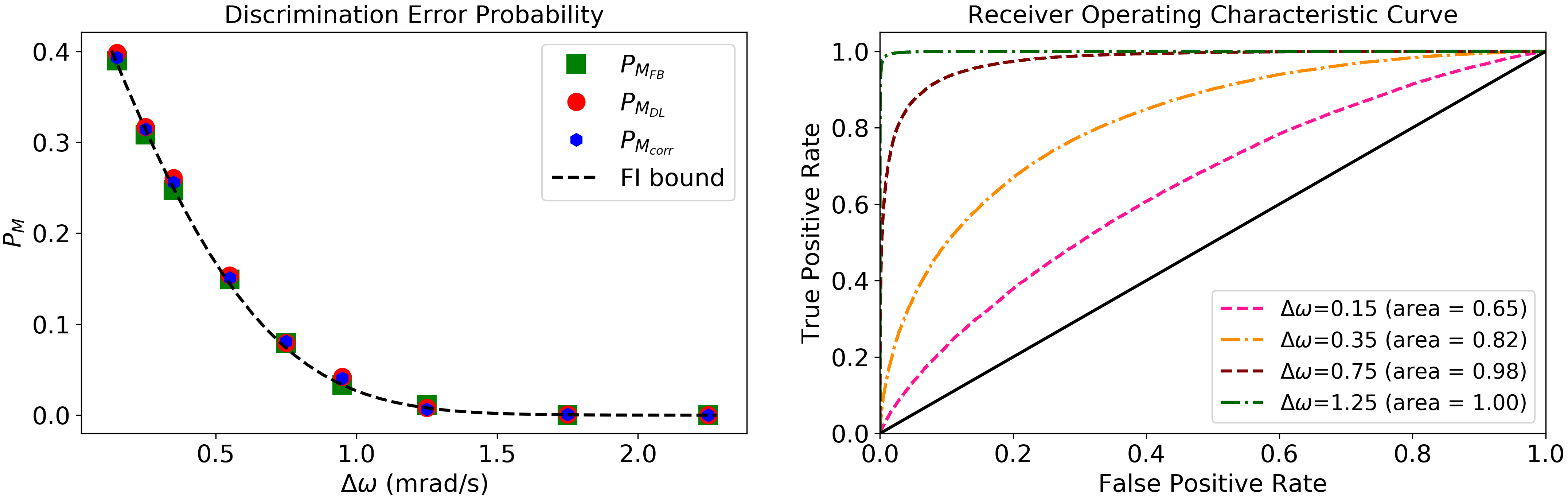}
\end{center}
\caption{Performance in the ideal model scenario. Left: discrimination error probabilities (Eq. (\ref{eq_error})) as a function of the frequency difference, $\Delta \omega.$ Full Bayesian, $P_{M_{FB}}$ (green squares), Deep Learning, $P_{M_{DL}}$ (red circles), correlations, $P_{M_{corr}}$ (blue hexagons), and analytical bound on $P_{M_{FB}}$ (dashed black). The input data were generated according to  Eq. (\ref{probideal})  with $g_1 = g_2 = \omega_1  = 10/(2 \pi)$ Hz, $\omega_2 = \omega_1 + \Delta \omega$, $\Delta t = 0.5$ sec, and a total measurement time of $T_{tot} = 500$ sec ($1000$ measurements). Right: receiver operating characteristic (ROC) curve and area under the curve (AUC) of $M_{DL}$ for different values of  $\Delta \omega,$ corresponding to the first, third, fifth and seventh points from left in the left figure.}
\label{error_freq_diff}
\end{figure}

\begin{figure*}[t!]
\begin{center}
\includegraphics[width=1\textwidth]{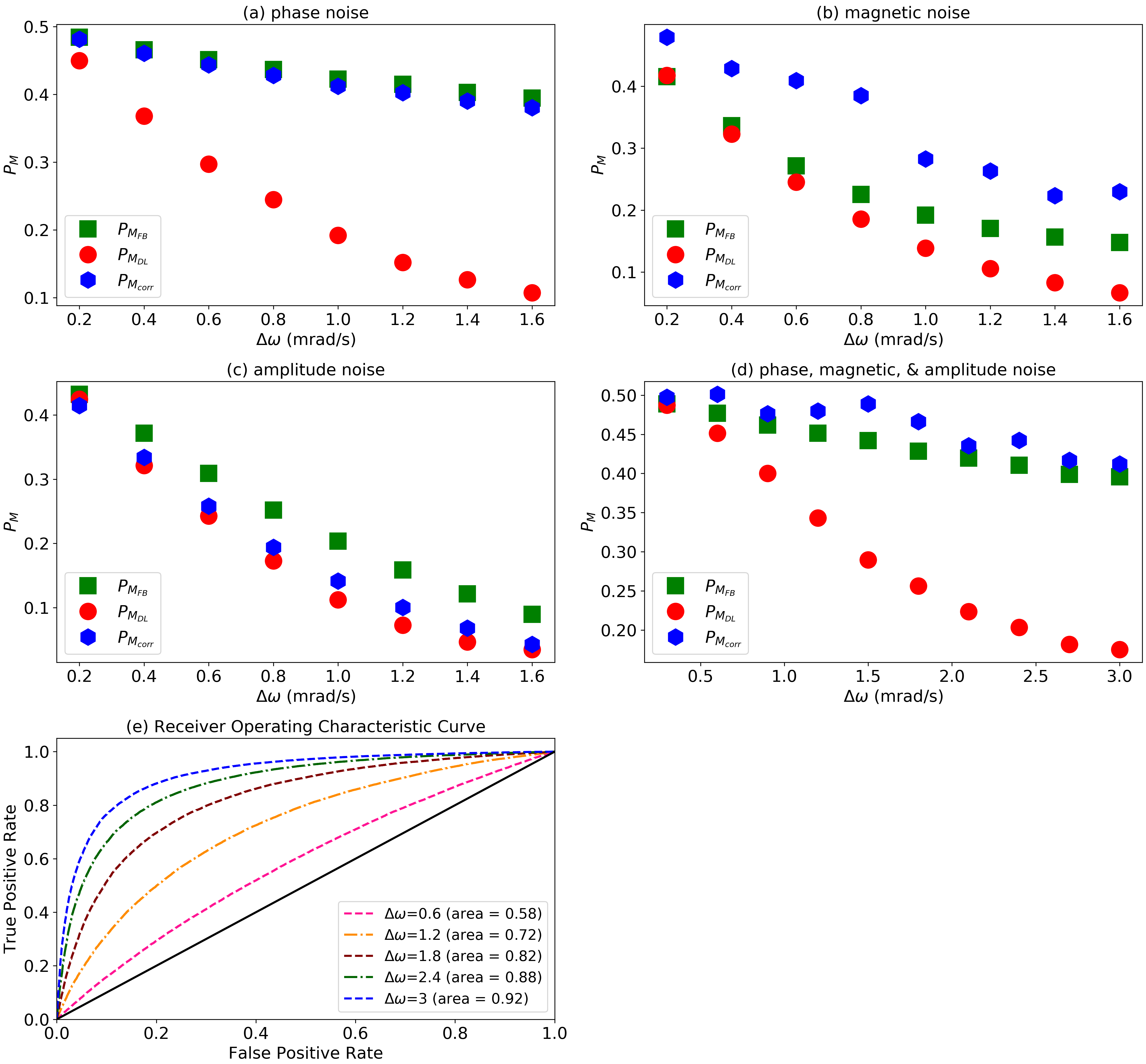}
\end{center}
\caption{Performance in noisy scenarios. Discrimination error probabilities (Eq. (\ref{eq_error})) as a function of the frequency difference, $\Delta \omega.$ Full Bayesian, $P_{M_{FB}}$ (green squares), Deep Learning, $P_{M_{DL}}$ (red circles), and correlations, $P_{M_{corr}}$ (blue hexagons). (a) phase noise - the random phase of the signal is randomly changed once during a single experiment at a random time interval. (b) magnetic noise - the probe is subject to a random magnetic field, which is randomly changed once during a single experiment at a random time interval. (c) amplitude noise - the amplitude of the signal has a different (random) value in each time interval.  (d) Mixed noise scenario, which includes all of the above noise models. See text for more details.  (e) ROC curve and AUC of $M_{DL}$ for different values of  $\Delta \omega,$ corresponding to the second, fourth, sixth, eighth and tenth  points from left in figure (d).
}
\label{error_freq_diff_noise}
\end{figure*}

\subsection*{Deep Learning method}
To overcome the model's lack of knowledge, we suggest using a supervised DL model, which we denote by $M_{DL}$. In particular, we consider a feed-forward fully connected neural network. The main reason for choosing a fully connected neural network is that the signal (frequency) information is encoded in the correlations between the values of different input neurons (measurement results), and in particular far apart input neurons. Similar to $M_{corr}$, we use a training dataset of measurement results of known signals (known labels) to train $M_{DL}$. We denote the labels of the two frequencies by $0$ and $1$. $M_{DL}$ is then applied to a test dataset  and results in estimations of the frequencies of the test measurement results.  Our DL model is a feed-forward fully connected neural network of four layers (two hidden layers) as depicted in Fig. (\ref{network}). While two hidden layers are sufficient for the scenarios considered in this work, it may be the case that more complex noise models would require to employ deeper neural networks. 
The first layer is called the input layer. The neurons of the input layer output the input data; in our case, the measurement results $x$ of a single experiment, to the second layer. The output of neuron $j$ in the second (hidden) layer is given by $f_j (z) = f(\sum_i w_{ij} x_i + b_j)$, where $f$ is the activation function, and $w_{ij}$ and $b_j$ are the weights and biases respectively. For the hidden layers we use the rectified linear (ReLU) activation function, $f(z) = \max(0,z)$. The output of the second layer is then fed as an input to the third layer and so on. The last layer is called the output layer. In our model the output layer has one neuron whose low and high activation levels are associated with the two possible labels (frequencies). For the output neuron we use the Sigmoid activation function. We use the mean squared error (MSE) between the outputs of the learning model and the labels of the train set as the loss function that is minimized during the training by optimizing the weights and biases of the model. Please note that there is no special reason for choosing the Sigmoid activation function with the MSE loss function; the softmax activation function together with the cross-entropy loss functions may be used as well. Overfitting is avoided by restricting the total numbers of nodes in the network (and hence, the number of free variables). In particular, for the examples considered in this work we use a second layer of $20$ nodes, and a third layer of $35$ nodes (a small modification of the number of neurons in each of the two hidden layers would not change the model's accuracy significantly). Regarding the test dataset, after the application of the Sigmoid activation function on the output of $M_{DL}$, we label the output by $1$ or $0$ according to whether it is $>0.5$ or $<0.5$ respectively. We then calculate $P_{M_{DL}}$ by the loss function (the MSE) between the output labels and the true labels.

\subsection*{Numerical analysis}
As a way of testing the performance of $M_{DL}$ in terms of frequency discrimination, we constructed numerical sets of measurement results, $x$, according to Eq. (\ref{probideal}) for two different frequencies, where the phase, $\phi_i$, was chosen randomly (uniformly distributed between $0$ and $2 \pi$) for each $x$. 
The input data were generated with $g_1 = g_2 = \omega_1  = 10/(2 \pi)$ Hz, $\omega_2 = \omega_1 + \Delta \omega$, $\Delta t = 0.5$ sec, and a total measurement time of $T_{tot} = 500$ sec ($1000$ measurements).
Part of the datasets were used for training and the remainder was used for testing the learning model. We compared the performance of $M_{FB}$ to the performance of $M_{DL}$ and $M_{corr}$. In Fig. (\ref{error_freq_diff}) we show the discrimination error probabilities, $P_{M_{FB}}$, $P_{M_{DL}}$, and $P_{M_{corr}}$ as a function of the frequency difference, $\Delta \omega$, between the two signals, as well as the corresponding $M_{DL}$ receiver operating characteristic (ROC) curves and areas under the curve (AUC). We considered a first layer of $1000$ nodes ($1000$ measurements), a second layer of $20$ nodes, and a third layer of $35$ nodes. This choice of number of nodes limits the free variable space and allows us to avoid overfitting without resorting to regularization methods.
In this ideal scenario, both $M_{corr}$ and $M_{DL}$ approach the optimal performance of $M_{FB}$ even though both methods have no a priori information on the physical model.

In order to provide indications on the performance of  $M_{DL}$ in real-world noisy scenarios we further considered a few more noise models and assumed that these noise models are ``unknown"   and hence, they are not taken into account in the Bayesian methods  $M_{FB}$ and  $M_{corr}$, which remain unchanged as described above. This serves as an indication  on how much better  the performance of  $M_{DL}$ could be in comparison to $M_{FB}$ and  $M_{corr}$ in a real-world scenario when the noise model is truly unknown to some extent.  The first noise model is still a phase noise. While previously we considered that the random (uniformly distributed) phase of the signal is constant during a single experiment, here we consider a scenario in which the random phase is changed once during a single experiment, where the second random phase is also uniformly distributed. Moreover, the time interval in which the phase change occurs is also uniformly distributed between the time intervals of a single experiment ($1000$ time intervals).   The discrimination error probabilities, $P_{M_{FB}}$, $P_{M_{DL}}$, and $P_{M_{corr}}$ as a function of the frequency difference, $\Delta \omega$, between the two signals are shown in Fig. (\ref{error_freq_diff_noise} (a)). It is clear that while the phase noise damages the discrimination capability of $M_{FB}$ and  $M_{corr}$, $M_{DL}$ is capable of learning the noise model.  The second noise model considers a magnetic noise $\delta b$, which modifies the Hamiltonian of the probe, Eq. (\ref{eq_single_signal}) to 
\begin{equation}
\label{eq_single_signal_mag_noise}
H_{s_i} = g_i \cos(\omega_i t + \phi_i) S_z + \delta b S_z.
\end{equation}
Similar to the phase noise, we assume that  $\delta b$ is changed once during a single experiment and that the time interval in which the change of $\delta b$ occurs is uniformly distributed between the time intervals of a single experiment. Each of the two values of $\delta b$ is Normally distributed with a zero mean and a standard deviation of $\sigma = g_i/5=2/(2 \pi)$ Hz.  The discrimination error probabilities, $P_{M_{FB}}$, $P_{M_{DL}}$, and $P_{M_{corr}}$ as a function of the frequency difference, $\Delta \omega$, between the two signals are shown in Fig. (\ref{error_freq_diff_noise} (b)). In this case $M_{DL}$ handles the magnetic noise better that  $M_{FB}$ and much better than $M_{corr}$. In the third noise model we consider noise in the amplitude of the signal. Specifically, we assume that the amplitude value is different in each time interval and that it is Normally distributed with a  mean of $g=10/(2 \pi)$ Hz (the previous value of the non-noisy amplitude) and a standard deviation that is equal to the mean value, that is, $\sigma = g = 10/(2 \pi)$ Hz. The discrimination error probabilities, $P_{M_{FB}}$, $P_{M_{DL}}$, and $P_{M_{corr}}$ as a function of the frequency difference, $\Delta \omega$, between the two signals are shown in Fig. (\ref{error_freq_diff_noise} (c)). In this case $M_{DL}$ performs slightly better than $M_{corr}$ and better than $M_{FB}$. Lastly, we consider the mixed-noise scenario where all of the above three noise models are includes. The discrimination error probabilities, $P_{M_{FB}}$, $P_{M_{DL}}$, and $P_{M_{corr}}$ as a function of the frequency difference, $\Delta \omega$, between the two signals are shown in Fig. (\ref{error_freq_diff_noise} (d)) and the  corresponding $M_{DL}$ ROC curves and AUC are shown in  Fig. (\ref{error_freq_diff_noise} (e)). It is  apparent that $M_{DL}$ is still  capable of learning the noise model while the performance of $M_{FB}$ and $M_{corr}$ is severely degraded when assuming that we have no further knowledge on the noise model. Of course, in case that we have more knowledge on the noise model, we may be able to modify the Bayesian methods accordingly. However, the implication of such a modification is that the optimization is performed with respect to a larger set of free variables, and therefore implies longer run times while the DL run time remains unchanged. Moreover, the above results suggest that Bayesian method could be very sensitive to the noise model; a minor unknown difference between the true noise model and the assumed noise model could result in a significantly reduced performance of the Bayesian method (say, for example, that there are three phase changes in a single experiment instead of two).

\begin{figure}[t]
\begin{center}
\includegraphics[width=1\textwidth]{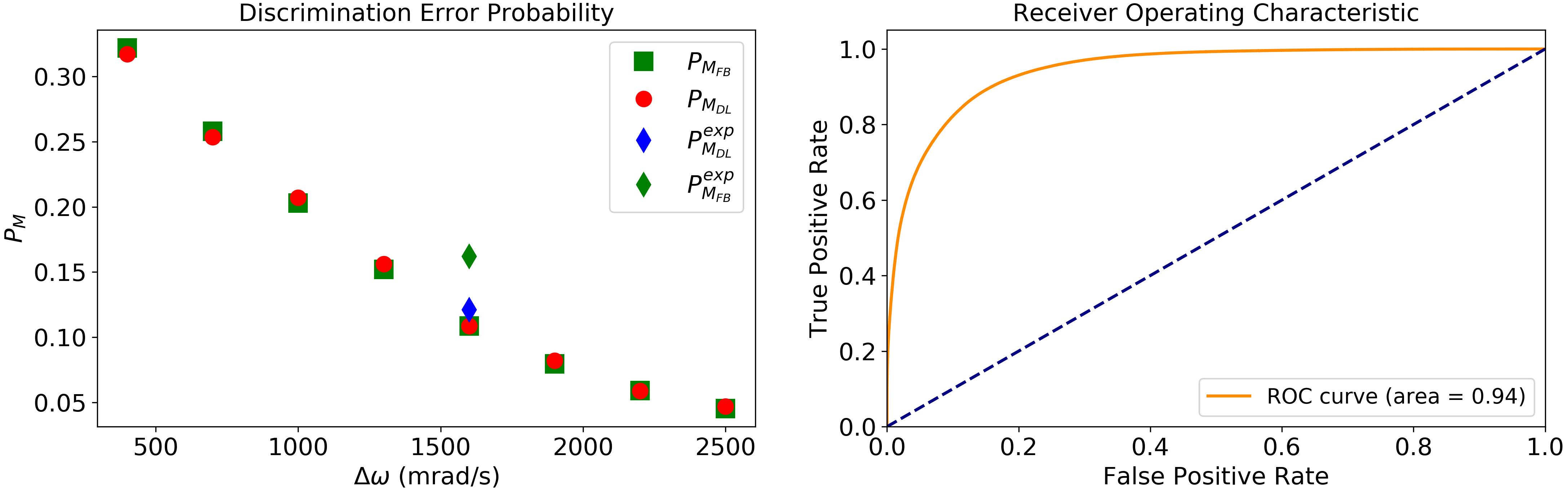}
\end{center}
\caption{Performance in the low-efficiency model scenario. Left: discrimination error probabilities (Eq. (\ref{eq_error})). Full Bayesian, $P_{M_{FB}}$ (green squares) and Deep Learning, $P_{M_{DL}}$ (red circles) on numeric data, Full Bayesian, $P^{exp}_{M_{FB}}$ (green diamond), and Deep Learning, $P^{exp}_{M_{DL}}$ (blue diamond)  on the experimental data, as function of the frequency difference, $\Delta \omega.$ The input numeric data were produced according to  Eq. (\ref{probloweff})  with $g_1=12.5$ KHz, $g_2=11.25$ KHz, $\omega_1=250$ Hz ,  
$\omega_2 = \omega_1 + \Delta \omega$, $\Delta t = 10\,\mu $sec and a total measurement time of $T_{tot} = 0.25$ sec ($25000$ measurements). Right: ROC curve and AUC of $M_{DL}$ on the experimental data, corresponding to the blue diamond  in the left figure.}
\label{error_freq_diff2}
\end{figure}

\subsection*{Experimental verification}
The NV center in diamond \cite{jelezko2006single,schirhagl2014nitrogen,doherty2013nitrogen} is one of the leading quantum probe systems for sensing, imaging and spectroscopy. Here we considered  frequency discrimination of measurement results obtained by a single NV center in ambient conditions. Two artificial signals were produced by a signal generator with frequencies $\omega_1 = 2\pi\times 250$ Hz and $\omega_2 =2\pi\times 251.6$ Hz. Each signal was measured for a total measurement time of $T_{tot}=220$ sec, with a time interval of $\Delta t = 10$ $\mu$s. From the row data, we generated strings of $25000$ measurement results ($T_{tot}=0.25$ sec) such that the phase corresponding to each $x$ can be considered as a random phase (no phase relation), and the frequencies cannot be resolved by a Fourier-Transform (see  Fig. \ref{two_freq} (left)). The low photon-detection efficiency of a true detection  ($m_s = 0$) and a false detection ($m_s = -1$) was $\sim 7.4 \%$ and $\sim 5.2 \%$ respectively, indicating low SNR and contrast.

In order to achieve a theoretical bound on the discrimination error, we considered  a theoretical model with a modified probability for a successful measurement, which is given by
\begin{equation}
Q(t) =\eta_{true} P(t) + \eta_{false}\left [1- P(t)\right],
\label{probloweff}
\end{equation}
where $P(t)$ is given by  Eq. (\ref{probideal}), and $\eta_{true}$ and $\eta_{false}$ are the true and false detection efficiencies respectively. Assuming that $\eta_{false}=0.7\eta_{true}$, we constructed numerical datasets according to Eq. (\ref{probloweff}), and set the amplitudes of the signals, $g_1$ and $g_2$, and the efficiency $\eta_{true}$ for each signal to match the experimental results according to two constraints: (i) The power spectrum at the frequency of the signal of the numerical data was required to be approximately equal to the power spectrum of the experimental data. (ii) The average of the experimental and numeric signals fulfilled  $\langle x \rangle=\frac{\eta_{true}+\eta_{false}}{2}$. For the numerical model we achieved  $P_{M_{FB}}\approx 10.8 \%$ and $P_{M_{DL}}\approx 11.6 \%$, (see Fig. \ref{error_freq_diff2} (left), green square and red circle under the diamonds). These results are consistent with the experimental data, for which we obtained $P^{exp}_{M_{DL}}\approx 12.1\%$ (Fig. \ref{error_freq_diff2} (left) blue diamond), reaching $P_{M_{FB}}$ without having any information on the model. Moreover, the Full Bayesian method on the experimental data obtained only $P^{exp}_{M_{FB}}\approx 16.2\% $ (Fig. \ref{error_freq_diff2} (left) green diamond). 
This difference is due to the fact that the experimental statistics differ slightly from our  probability function; while for the Bayesian method this creates a problem, the DL method is able to learn this difference and take it into account. This difference is expected to be much more dramatic in real nano-NMR experiments in which there are much more uncertainties of the model.
In addition, we analyzed $P_{M_{FB}}$ and $P_{M_{DL}}$ on the numerical data as a function of the frequency difference, $\Delta \omega$. The results are shown in Fig. (\ref{error_freq_diff2} (left)). The ROC curve and AUC  of $M_{DL}$  on the experimental data are shown in Fig. (\ref{error_freq_diff2} (right)) .

It is worth noting that due to the relatively large window size of $25000$, a full analysis of $M_{corr}$ is not possible within a reasonable time scale on a common computer. 
Partial analysis (taking into account segments of two-point correlations only) of $M_{corr}$ of both the numerical model and the experimental data yielded   $P_{M_{corr}}\gtrsim 0.4$.   
This indicates that DL could indeed be the better choice when there is a lack of knowledge on the model.

\subsection*{Comparison to other Machine Learning methods} 
So far we have shown that DL methods are useful for the problem of frequency discrimination in the nano-NMR settings. In this section we ask whether other machine learning methods could be useful for this task and if so, how these methods perform compared to DL. 

Any method that is able to discriminate between two signals of near-by frequencies, as we have considered in previous sections, should be able to learn and acquire the information on the signals from the correlations between different measurement results (different $x_i$). Hence, any successful discrimination method should involve some non-linearity. Indeed, a fully connected neural network with only linear layers fails in the considered discrimination problem (the achieved error probability is $1/2$). We tested the performance of three other linear learning methods, namely, logistic regression (with no interaction terms), K nearest neighbours and supported vector machines (SVM) with a linear kernel, in the ideal model scenario  (Fig. \ref{error_freq_diff}). Similarly to a fully connected linear neural network, these  methods completely fail to discriminate between the signals and achieve an error probability of $1/2$ for all values of $\Delta \omega.$ 

\begin{figure}[t]
\begin{center}
\includegraphics[width=1\textwidth]{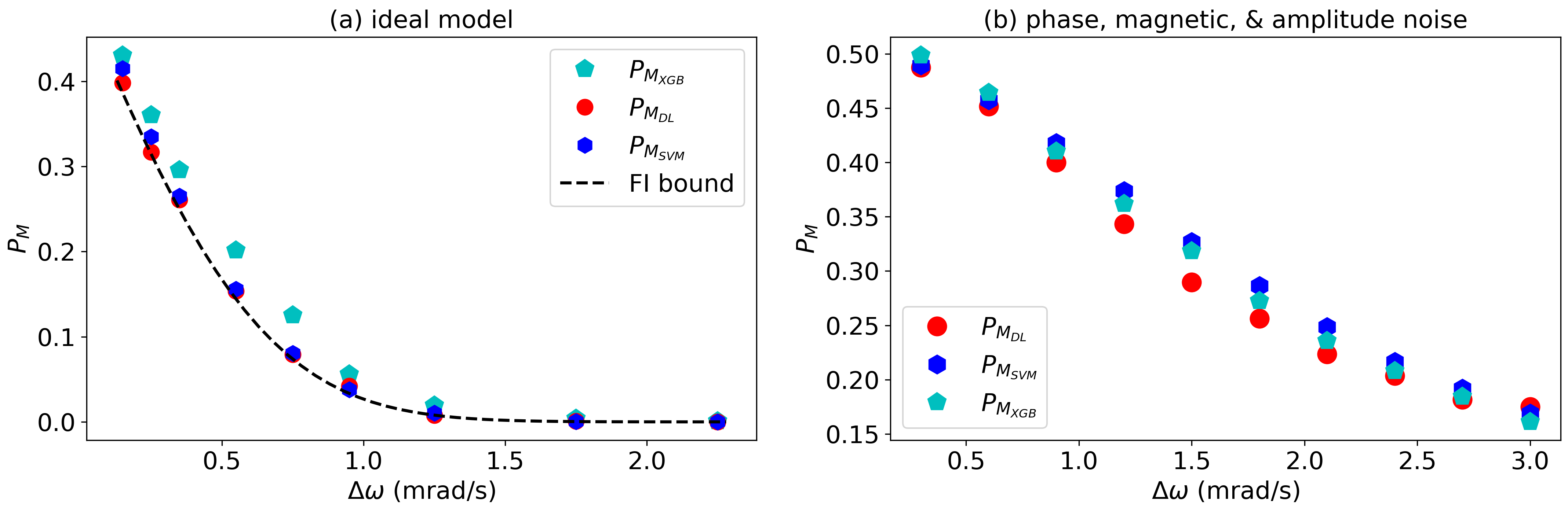}
\end{center}
\caption{Comparison to other Machine Learning methods. Discrimination error probabilities (Eq. (\ref{eq_error})) of Deep Learning, $P_{M_{DL}}$ (red circles), Supported Vector Machines, $P_{M_{SVM}}$ (blue hexagons), and XGBoost, $P_{M_{XGB}}$ (magenta pentagon).  (a) Ideal model (see Fig. \ref{error_freq_diff}). (b) Mixed noise model (see Fig. \ref{error_freq_diff_noise} (d)).}
\label{ML_compar}
\end{figure}

Regarding non-linear models, we considered two models, SVM with the non-linear radial basis function (rbf) kernel and XGboost, which is an implementation of gradient boosted decision trees (in our case we consider non-linear boosting as linear boosting fails). We tested these two models in the ideal model  scenario (Fig. \ref{error_freq_diff}) as well as in the mixed noise model scenario (see Fig. \ref{error_freq_diff_noise} (d)).
The results are shown in Fig. \ref{ML_compar}. It can be seen that these two methods achieve accuracies (discrimination error probabilities) which are very similar to the accuracies obtained by DL. However, there is a big difference in terms of required computational resources as these two methods consume more memory compared to DL, and require much longer running times of, for example,  $\sim20$ hours compared to $\sim20$ minutes by DL. Indeed, it is not feasible to use these methods for the discrimination in the case of the experimental data, where the size of the inputs is much larger (input strings of $25000$ compared to input strings of $1000$). While we have not made an exhaustive study and analysis of machine learning methods, which is beyond  the scope of this work, these findings strengthen the possible advantage and benefit of DL method for data processing of nano-NMR experimental results. 

\section*{Frequency resolution}

\begin{figure}[t]
\begin{center}
\includegraphics[width=0.6\textwidth]{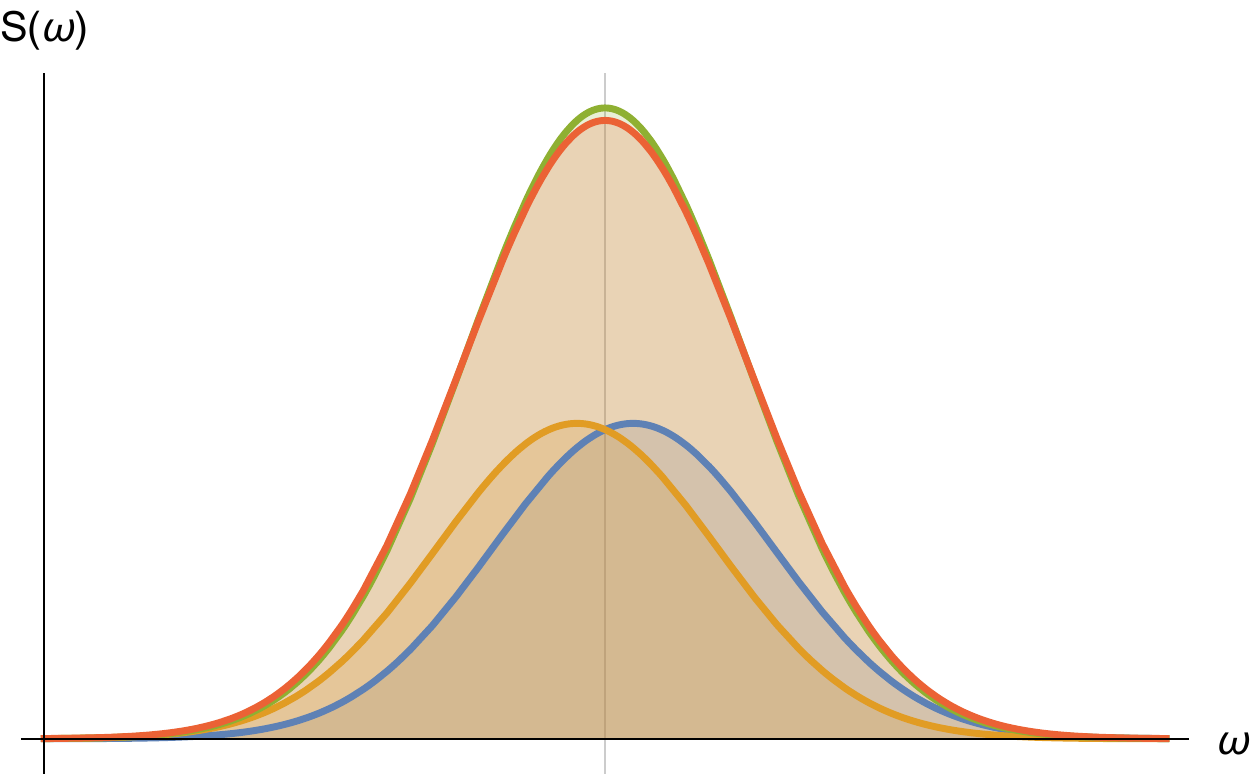}
\end{center}
\caption{The problem of frequency resolution. The observed signal could be one of two possible signals that should be resolved. One signal (red) has two near by frequencies (blue and orange) and corresponds to their sum. The second signal  (green) has one frequency, which is centered between the two near by frequencies of the first signal. The closer the two frequencies are, the harder it is to resolve between the two signals.
}
\label{resolution}
\end{figure}

In this section we considered the problem of discrimination between a signal with a single frequency and a signal with two proximal frequencies centred at the value of the single frequency (Fig. \ref{resolution}). We assumed that the signals have strong amplitude and phase noise, which we model by the Ornstein-Uhlenbeck (OU) process, motivated by NV probed statistically polarized nano-NMR experiments \cite{staudacher2015probing,mamin2013nanoscale,staudacher2013nuclear}. The OU process is a stochastic random process, which is a stationary, a Gaussian and a Markov process. It is given by $x(t+dt)=x(t)-\frac{1}{\tau}x(t)dt+\sqrt{c}dW(t)$, where $\tau$ and $c$ are positive constants called, respectively, the relaxation time and the diffusion constant, and  $dW(t)$ is a temporally uncorrelated normal random variable with mean $0$ and variance $dt$. The environmental noise experienced by an NV center is faithfully  modelled by an OU process \cite{Lange}.

Specifically, the Hamiltonian of the probe is given by 
\begin{equation}
H =  \left( \sum_{i=1}^n A_i\left(t\right) \cos[\delta_i t ] - B_i\left(t\right) \sin[\delta_i t ]\right) S_z, 
\end{equation}
where $A_i$ and $B_i$ undergo an OU process due to the amplitude and phase noise, and $\delta_i$ are the frequencies. The probability for a successful measurement (one) is 
\begin{eqnarray}
P(t)&=&\sin\left[\sum_{i=1}^n\frac{A_i\left(t\right)}{\delta_i}(\sin[\delta_i t ]-\sin[\delta_i (t-\Delta t)])\right. \nonumber\\
&+& \left.\frac{B_i\left(t\right)}{\delta_i}(\cos[\delta_i t ]-\cos[\delta_i (t-\Delta t)])+\frac{\pi}{4}\right]^2,
\label{probOU}
\end{eqnarray}
where $n=2$ and $\delta_i=\delta_c\pm\Delta/2$. For two frequencies $\Delta$ is finite, and for a single frequency $\Delta=0$.

\subsection*{Numerical analysis} 
We constructed numerical datasets according to Eq. (\ref{probOU}) where $A_i(t)$ and $B_i(t)$ follow OU processes with mean $\mu = 0$, volatility $\sigma = \frac{\pi}{10}\sqrt{\frac{4}{\pi T_2}}$, and reversion speed $\theta=1/T_2$, where $T_2=256$ sec is the coherence time of the signal. In addition, we fixed  $T_{tot}=2 T_2$ and $\Delta t=1$ sec. 
We tested the performance of $M_{DL}$ as a function of the frequency difference, $\Delta$, in comparison to $M_{FB}$ and $M_{corr}$. In $M_{FB}$ the maximal log-likelihood was calculated over the random OU processes. For each string of measurement results $x$, we considered the single frequency signal with $\Delta=\Delta_0=0$ and the signal of two near-by frequencies with $\Delta=\Delta_n>0$, where $\Delta_n$ corresponds to the numerical value of the frequency difference between the two frequencies. We generated many sets of random OU processes denoted by $O_k$ and calculated  
\begin{equation}
L_1 = \max_{O_k} L(O_k|x,\Delta_0),\quad  L_2 = \max_{O_k} L(O_k|x,\Delta_n),
\end{equation}
where 
\begin{eqnarray}
 L(O_k|x,\Delta_i) &=& \sum_j (x_j \log P(t_j,\Delta_i,O_k)\nonumber\\
  &+& (1-x_j) \log (1- P(t_j,\Delta_i,O_k))).
\end{eqnarray}
We estimated the signal as a single frequency signal or as a signal of two frequencies according to the larger likelihood; that is
\begin{equation}
\Delta_{est} = \begin{cases}
 \Delta_0 & L_1 > L_2\\
\Delta_n &\text{otherwise}.
\end{cases}
\end{equation}

\begin{figure}[t!]
\begin{center}
\includegraphics[width=1\textwidth]{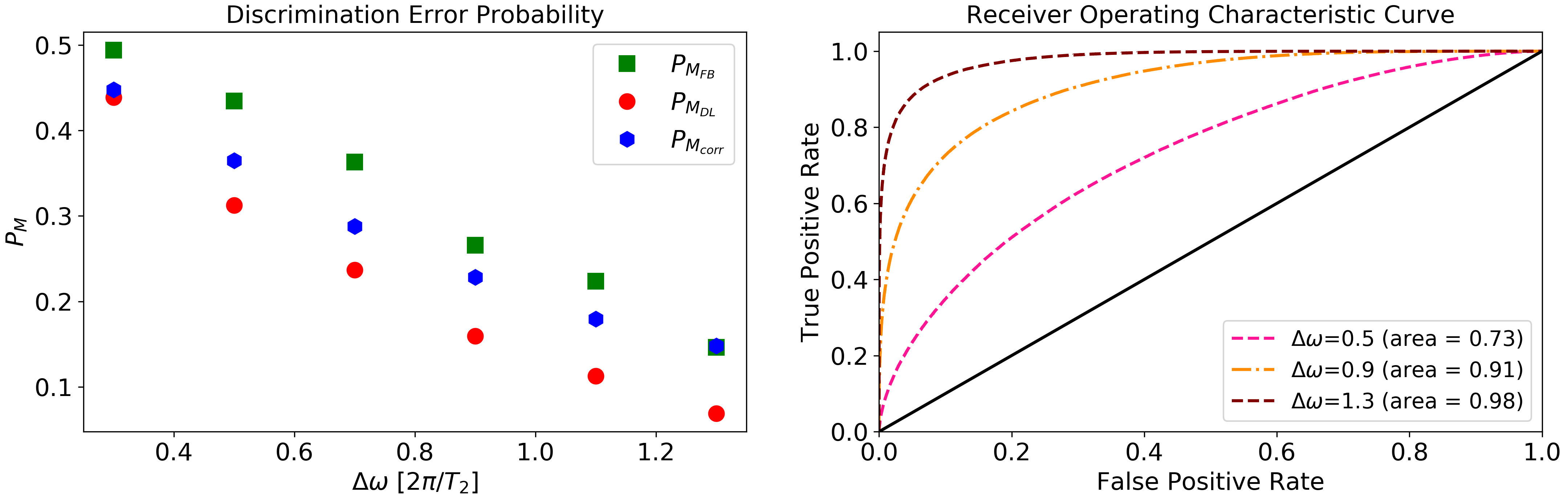}
\end{center}
\caption{Performance in the noisy frequency resolution scenario. Left: discrimination error probabilities (Eq. (\ref{eq_error})). Full Bayesian, $P_{M_{FB}}$ (green squares), Deep Learning, $P_{M_{DL}}$ (red circles), and  correlations, $P_{M_{corr}}$ (blue hexagons), as a function of the frequency difference, $\Delta \omega$. The input data were produced according to  eq. (\ref{probOU}) with $T_{tot} = 2 T_2$. Right: ROC curve and AUC of $M_{DL}$ for different values of  $\Delta \omega,$ corresponding to the second, fourth, and sixth points from left in the left figure.}
\label{error_freq_diff3}
\end{figure}

Fig.  \ref{error_freq_diff3} (left) shows the error probability as a function of the frequency difference. The $M_{DL}$ results were better than the results of $M_{corr}$ as well as the results of $M_{FB}$. Interestingly, even though  $M_{FB}$ has full knowledge of the noise model it achieves a larger error probability than $M_{DL}$. We note that increasing the number of OU processes, $O_k$, in the above likelihood calculation does not improve  $P_{M_{FB}}$.  While $M_{DL}$ and $M_{corr}$ could reach a result within $\sim 45$ min, $M_{FB}$ did so within $\sim 7$ hours (CPU times, both considered on the same common PC without utilizing GPU). The $M_{DL}$ ROC curves and AUC are shown in  Fig. \ref{error_freq_diff3} (right).
These numerical results provide a strong indication that DL methods can potentially identify molecules based on their NMR signal extremely fast, which may be a useful tool in probing chemical reactions at the nano scale. 

\subsection*{Theoretical implications} 

While the numerical advantages of machine learning methods  were already shown \cite{santagati2018magnetic,granade2012robust}, their theoretical value was not demonstrated before. Beyond the practical interest of utilizing machine learning methods in the nano-NMR frequency resolution problem, machine learning methods, and in particular DL methods, could also have a considerable theoretical value.

Generally in estimation problems, the MSE of an estimator $M$  for a given unseen test input $x$ can be written as 
\begin{equation}
\mathrm{E}\left[\big(y-M\left(x\right)\big)^2\right]=\Big(\mathrm{Bias}\left[M(x)\right]\Big)^2 + \mathrm{Var}\left[M(x)\right] +  \mathrm{Var}\left[\epsilon\right],
\label{MSE} 
\end{equation}
where $y$ is the true label, $M(x)$ is the estimated label, $\mathrm{E}$ is the expectation value with respect to the training set, the bias of $M$ is given by $\mathrm{Bias}\left[M(x)\right]=\mathrm{E}\left[M(x)\right]-M(x)$, the variance of $M$ is given by $\mathrm{Var}\left[M(x)\right]=\mathrm{E}\left[M(x)^2\right]-\mathrm{E}\left[M(x)\right]^2$, and $\mathrm{Var}\left[\epsilon\right]$ is the  irreducible error due to the (zero mean) noise $\epsilon$. The error probability, $P_M$, is then obtained by the expectation value of the MSE, $\mathrm{E}\left[\big(y-M\left(x\right)\big)^2\right]$, with respect to the test set. 

An unbiased estimator is an estimator $M$ for which we have that $\mathrm{Bias}\left[M(x)\right]=0$. An optimal unbiased estimator has a minimal variance, which is known as the minimum variance unbiased (MVU) estimator. However, from Eq. (\ref{MSE}) it is seen that an MVU estimator is not necessarily an optimal estimator which minimizes the MSE. Indeed, it is known that biased methods can outperform the unbiased ones \cite{efron1975biased,eldar2008rethinking,james1992estimation}. In this case the magnitude of the bias is increased and $\Big(\mathrm{Bias}\left[M(x)\right]\Big)^2>0$, but the variance $\mathrm{Var}\left[M(x)\right]$ is significanlty decreased such that the MSE is smaller than the MSE of an MVU estimator.      Such strategies of error reduction are used ubiquitously in image restoration  \cite{demoment1989image,meng2004modified} and beamforming applications \cite{cox1987robust,carlson1988covariance}. Moreover, it is known that biased methods can be superior in various spectral analysis applications \cite{stoica1997introduction}.
Despite its superiority there are only a few structured methods in which such a biased estimator can be constructed \cite{eldar2008rethinking} and in most cases the search for such  estimators is extremely challenging, especially as it is unknown if such an estimator exists.  

Our numerical analysis of $M_{FB}$ has converged to the final result, however, the method has resolved the two frequencies with a higher error rate than the error rate of $M_{DL}$. Since $M_{FB}$ is an MVU estimator, our results indicate that for the model at hand the unbiased full model Bayesian analysis is not optimal, and that a superior biased method exists. This brings up an extra advantage of DL as the search for a biased method in usually done in an ad-hc manner. Moreover, in most cases there is no way of knowing if a superior method to the unbiased method exists.  
Hence, our results provide some hope that DL methods could be used as an analytical tool for identifying superior estimators, and in particular, for identifying the ultimate limits of resolution problems.
 
\section*{Conclusion} 
In conclusion, we showed that DL methods are able to mitigate the effect of the inherent strong noise in the nano-NMR settings.
In particular, the DL neural networks effectively learn the noise model, even when no prior knowledge on the noise model is assumed. This is a crucial property of the DL methods as in many realistic nano-NMR scenarios the noise model is complex and not accurately known. We investigated the performance of DL methods in the problems of frequency discrimination and frequency resolution. We showed that DL methods can outperform Bayesian methods when full knowledge of the noise model is not available  and that DL methods can analyze a test signal as accurately as numerically demanding Bayesian methods even though Bayesian methods have full knowledge of the noise model, and the DL methods have no prior knowledge at all.

DL methods can perform better than Bayesian methods when the noise model is not precisely known or when the noise model is known but it is a complex model. In the first case DL methods can achieve better results than Bayesian methods as  DL methods do not assume prior knowledge on the model while Bayesian methods rely on precise knowledge of the model. This was demonstrated in the case of frequency discrimination in the noisy scenario, as well as in the analysis of the experimental data. In the second case the results of both methods may be similar, but the consumption of computational resources of  Bayesian methods can be much larger compared to the resources required by DL methods, as was demonstrated in the problem of frequency resolution of noisy signals.   

Our results can be seen as a strong indication that DL methods will turn out to be the method of choice when analyzing spectroscopic nano-NMR data.  In addition, our results indicate that DL methods could be utilized as a tool that may enable to identify superior biased estimators and ultimate limits of resolution problems, which are otherwise difficult to obtain \cite{rotem2017limits}.

\section*{Acknowledgements} A.R, N.A  have received funding from the European Union's Horizon 2020 research and innovation programme under grant agreement No. 770929  ERC consolidator grant QRES and  the collaborative projects ASTERIQS and Hyperdiamond.

\section*{Competing Interests}
 The authors declare that they have no competing financial or non-financial interests.
 
\section*{Data availability}
The authors declare that all relevant data supporting the findings of this study are available within the paper (and its Supplementary information files).

\section*{Authors Contribution}
A. Retzker. and Z.R. supervised the project.
N.A., A.Retzker. and Z.R. performed the theoretical analysis.
N.A. performed the deep learning and machine learning analysis.
N.A. and A. Rotem carried out the spectral resolution numerical analysis.
L.P.M. and F.J. conducted the experiment.      
N.A. took the lead in writing the manuscript with support from A.Retzker. and Z.R..

\bibliography{Bib_Alex}

\end{document}